\documentclass[pre,twocolumn,showpacs,preprintnumbers,amsmath,amssymb,floatfix]{revtex4}
\usepackage{epsfig,enumerate}
\usepackage{graphicx}
\usepackage{bm}

\begin{document}

\title{Correlation entropy of synaptic input-output dynamics}
\author{Ingo C. Kleppe}
\altaffiliation[Present Address: ]{%
Department of Physiology, University College London, Gower Street,
London WC1E 6BT, U.K.}
\author{Hugh P.C. Robinson}
\affiliation{%
Department of Physiology, University of Cambridge, Downing Street,
Cambridge, CB2 3EG, U.K.}
\date{\today}

\begin{abstract}
\noindent The responses of synapses in the neocortex show highly
stochastic and nonlinear behavior. The microscopic dynamics
underlying this behavior, and its computational consequences during
natural patterns of synaptic input, are not explained by
conventional macroscopic models of deterministic ensemble mean
dynamics. Here, we introduce the correlation entropy of the
synaptic input-output map as a measure of synaptic reliability
which explicitly includes the microscopic dynamics. Applying this
to experimental data, we find that cortical synapses show a low-dimensional
chaos driven
by the natural input pattern.
\end{abstract}


\maketitle
\section{Introduction}
The excitatory cortical synapse is an example of a nonlinear
biological system with a high level of intrinsic noise. This leads
to a complex relationship between the input -- the times of
presynaptic action potentials (APs) -- and the output -- the
variable amplitudes of excitatory postsynaptic potentials (EPSPs).
The underlying mechanisms of this dynamics are not well
understood. However, cortical synaptic transmission has been
studied extensively by measuring the ensemble mean responses to
short stimulus trains. Such responses show short-term `plasticity'
or activity--dependent changes, both augmenting and depressing
\cite{tsodyks:1997, abbott:1997, varela:1997, dobrunz:1997,
thomson:1997b}, as do measurements of field potentials, which
represent the spatial average of  EPSPs in a large population of
synapses, all driven with the same timing \cite{dobrunz:1999}.

Current models of the dynamics of short-term
plasticity are mean-field approximations \cite{tsodyks:1997,
abbott:1997, varela:1997}, systems of deterministic differential
equations describing the average flux of transmitter between
different functional pools. In this treatment, the large `synaptic
noise' around the ensemble mean of individual responses is
considered to be extrinsic to the underlying dynamics of the
synaptic response. In fact, though, it is known that the
trajectory of individual responses contains a large amount of (at
least short-term) predictability or determinism which is lost by
the averaging of responses. For example, immediately following a
failure to release transmitter in response to a presynaptic spike,
there is a greatly raised probability of release at a
closely-subsequent spike, and vice versa (e.g.
\cite{stevens:1995}).

As a consequence, characterizing the input-output dynamics of the
synapse by the deterministic dynamics of the ensemble mean does not
capture the full predictability of the synapse. From a
neurobiological point of view, a useful goal in understanding the
operation of neural circuits is to estimate the rate at which
predictability is lost, or equivalently the rate of production of
new information by the dynamics of individual synapses, during
natural sequences of input. Our aim here is to do this in a way
which captures nonlinear correlations in the fluctuations around
ensemble mean responses, which embody much of the predictability of
synaptic transmission. In addition, this will shed light on the
dynamical nature of synaptic transmission. For example, do the noise
of the input and the intrinsic noise of the synapse result in
behavior like stochastic chaos \cite{rand:1991}? In this study, we
describe a general, nonlinear approach to this problem, using the
correlation entropy of the input-output map of the synapse.

\section{Correlation entropy estimated from input-output time series}

A sequence of synaptic responses to an arbitrarily-timed
spike-train input may be thought of as a map from input-output
history to the amplitude of the next event:
\begin{equation}
\mathbf{h_i} \mapsto A_{i+1}
\end{equation}
\noindent where
\begin{align}
\label{eqn:historyvector} {\bf h}_{i} & =[{\bf A}_i^m ; {\bf{\Delta
t}}_i^n] \nonumber \\
{} & =[A_{i-m+1},A_{i-m+2},.....,A_i,  \\
{} & \hspace{3ex} \Delta t_{i-n+1}, \Delta t_{i-n+2},..... \Delta
t_{i}] \nonumber
\end{align}
\noindent and $m$ is the number of amplitude dimensions, $n$ is
the number of interspike interval dimensions in the history,
$\mathbf{A}$ represents the amplitudes and $\mathbf{\Delta t}$ the
intervals preceding each response. Equivalently, $\mathbf{h_i}
\mapsto \mathbf{h_{i+1}}$ defines the dynamics in an input-output
time delay space (see \cite{cao:1998a, cao:1998b}).\\
The correlation entropy $K_2$ is a lower bound for the
Kolmogorov-Sinai entropy of a dynamical system
\cite{grassberger:1983}, which can be calculated from correlation
sums. We extend the definition given in \cite{grassberger:1983} to
this bivariate, input-output process, distinguishing input (time
intervals) and output (amplitude) dimensions. Let
\begin{equation}
\mu=\ln\frac{C(m,n,\delta,\epsilon)}{C(m+1,n+1,\delta,\epsilon)} -
\ln\frac{C(n,\delta)}{C(n+1,\delta)}.
\end{equation}
The first term is the correlation exponent for the joint
input--output process, while the second term is that for the input
process alone.
 \noindent Then the correlation entropy of the input--output process
\begin{equation}
K_2 = \lim_{m,n \rightarrow \infty}\lim_{\delta,\epsilon
\rightarrow 0 }\lim_{N \rightarrow \infty}
\mu(m,n,\delta,\epsilon)
\end{equation}
\noindent where $N$ is the number of data points. The correlation
sums are given by
\begin{align}
& C(m,n,\delta,\epsilon)  = \frac{2}{N(N-1)} \times\nonumber \\
&
\sum_{i=1}^N\!\sum_{j=i+1}^N\Theta(\epsilon\!-\|\mathbf{A}_i^m\!-\!\mathbf{A}
_j^m\|)\Theta(\delta-\!\|\mathbf{\Delta t}_i^n-\mathbf{\Delta
t}_j^n\|) \label{eqn:corrsum_io}\\
& C(n,\delta)  =\frac{2}{N(N-1)}
\sum_{i=1}^{N}\!\sum_{j=i+1}^N\Theta(\delta-\!\|\mathbf{\Delta
t}_i^n-\mathbf{\Delta t}_j^n\|) \label{eqn:corrsum_i}
\end{align}
\noindent where $\Theta$ is the Heaviside step function,
$\|\cdots\|$ denotes the maximum norm and $\epsilon$ and $\delta$
are the neighborhood extents in amplitude and interval dimensions
respectively. $K_2$ measures the uncertainty per synaptic event in
the postsynaptic potential from the entire dynamics of synaptic
transmission while subtracting the input uncertainty. It is useful
to examine the convergence behavior of $\mu$, the estimate of $K_2$,
as a function of the output neighborhood size $(\epsilon)$. As the
neighborhood shrinks, $\mu$ for stochastic processes approaches
infinity, or for chaotic deterministic processes, a finite positive
value \cite{gaspard:1993, schittenkopf:1997,faure:1998}. For small
\emph{extrinsic} noise, for example, it is in principle possible to
separate the entropy that is due to deterministic dynamics from the
noise, and to estimate the noise level as the neighborhood radius
below which $\mu$ starts to rise sharply.
\begin{figure}
\begin{center}
\epsfig{file=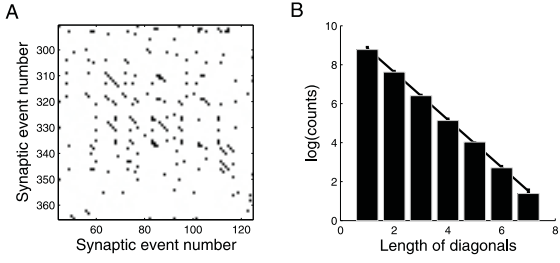,width=\linewidth}
\end{center}
\caption{Input-output recurrence plot of synaptic transmission data.
Stimulation parameters as in Fig.~\ref{fig:ensemble_data}B, D--E.
{\bf A}: Excerpt of a recurrence plot for neighborhood dimensions
$\epsilon=0.35$ mV, $\delta = 290$ms. Matrix elements $[i,j]$ are
colored black when
$\Theta(\epsilon-\|\mathbf{A}_i-\mathbf{A}_j\|)\Theta(\delta-\|\mathbf{\Delta
t}_i-\mathbf{\Delta t}_j\|) = 1$. $m = n = 2$. {\bf B}: Cumulative
logarithmic histogram of diagonal lengths in a recurrence plot of a
complete time series (30 mins). The correlation exponent of the
joint input--output process is estimated from the slope of this plot
(1.75).} \label{fig:recurrence_example}
\end{figure}
We used a method that estimates $\mu$ from the slopes of the
logarithmic distributions of diagonal lengths in recurrence plots
(RP)\cite{faure:1998}, rather than directly from
Eqns.~(\ref{eqn:corrsum_io},\ref{eqn:corrsum_i}). The RP is a
matrix representing similarity in local history between all pairs
of embedding points in a time series
(Fig.~\ref{fig:recurrence_example}A). This method is robust
against slow nonstationarity in the data \cite{webber:1994}, and
is also computationally efficient, since it requires only one
calculation of the distances between points at a low embedding
dimension. As expected, the distributions of diagonal lengths
could be fitted very well with an exponential, allowing
unambiguous measurement of the correlation exponents
(Fig.~\ref{fig:recurrence_example}B).

\section{The noise-driven logistic map}

In this section we illustrate the method using the logistic map
perturbed with noise in each iteration. This is given by
\begin{equation}
x_{i+1}=\|a(x_i+\xi_i)(1-x_i-\xi_i)\|\;\mathrm{mod}\; 1
\label{eqn:logisticnoise}
\end{equation}
Here the noise values $\xi_n$, drawn randomly from a Gaussian
distribution with a standard deviation $\sigma_\xi$, correspond to
the input time series $\Delta t_i$ above. Analogously the output
time series is given by the set of $x_n$, corresponding to the set
of $A_i$.
\begin{figure}
\begin{center}
\epsfig{file=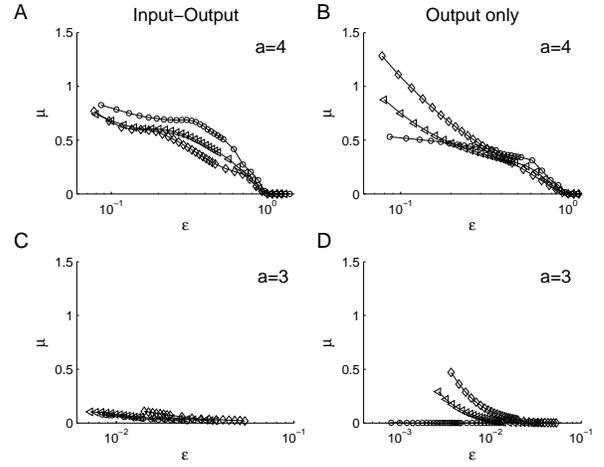,width=\linewidth}
\end{center}
\caption{Input-output correlation entropy of the noise-driven
logistic map (eqn.\/~\ref{eqn:logisticnoise}). Time series consisted
of 5000 points, and results from 500 trials were averaged. Symbols
denote various noise levels relative to the standard deviation of
the unperturbed map $\sigma_{\xi} /\sigma_{up}$: $\diamond = 0.5$,
$\lhd = 0.25$, $\circ = 0.01$, $m=n=6$. $x_0= 0.7$, $a=4$ (A,B) and
$a=3$ (C,D). {\bf A}: $\mu$ as a function of $\epsilon$ for $\delta
\rightarrow 0$ {\bf B:} $\mu$ as a function of $\epsilon$ estimated
solely from the output time series ($\delta \rightarrow \infty$).
{\bf C:} As A but for $a=3$ {\bf D:} As B but for $a=3$.}
\label{fig:logistic_noise}
\end{figure}
First, we chose $x_0=0.7$ and $a=4$, which produces a chaotic
process in the unperturbed case.

\begin{figure*}[t!]
\begin{center}
\epsfig{file=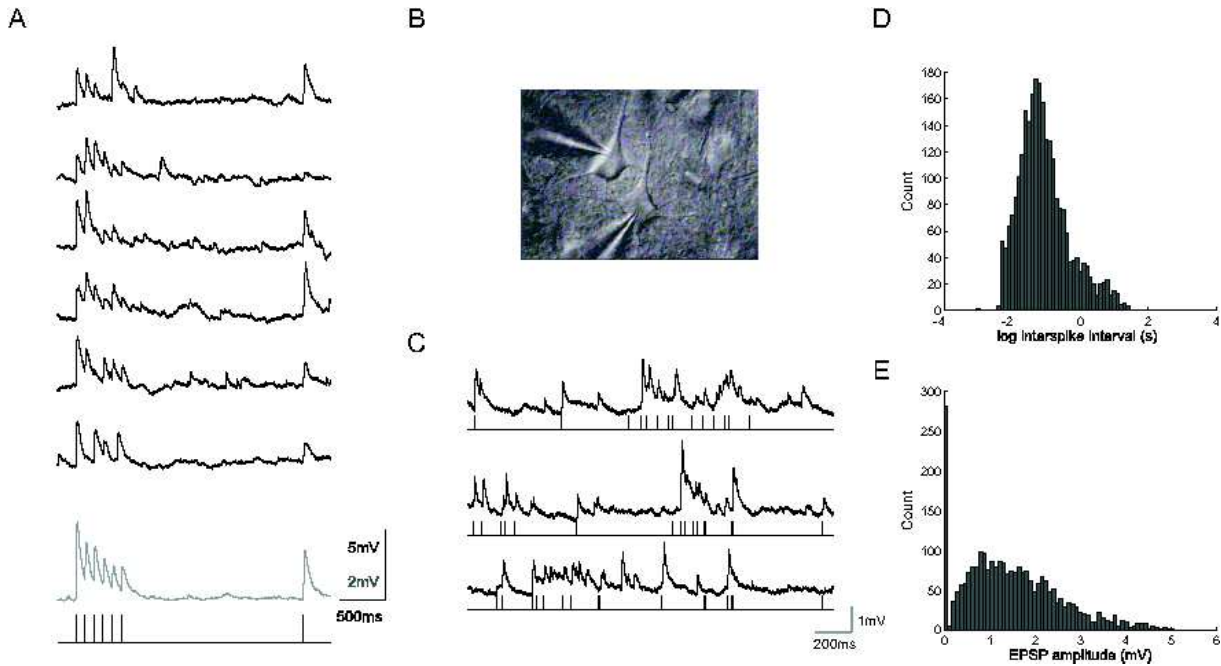,width=0.9\linewidth} \caption{Synaptic
transmission at a cortical synapse. {\bf A}: 6 consecutive
postsynaptic responses (top) and ensemble mean response (bottom, 50
responses) to the presynaptic stimulation pattern indicated below.
{\bf B:} Whole-cell recording in synaptically coupled pyramidal
neurons of the rat cortex. {\bf C}: A segment of a continuous
recording showing evoked and spontaneous EPSPs (upper traces) during
naturalistic stimulation (APs in lower traces) with parameters:
$R_p= 30$Hz $\lambda_{b}=0.2$Hz $\tau_{b}=200$ ms, average rate 1.2
Hz (see text). {\bf D, E}: Corresponding distributions of interspike
intervals (D) and postsynaptic EPSP amplitudes (E) for an experiment
lasting 30 minutes.} \label{fig:ensemble_data}
\end{center}
\end{figure*}
The profile of the convergence of $\mu$ as $\epsilon$, the output
neighbourhood radius, approaches zero is illustrated in
Fig.~\ref{fig:logistic_noise}, at three different strengths of the
driving noise $\sigma_{\xi}$. Fig.~\ref{fig:logistic_noise}A shows
the profile for the input-output correlation entropy. When $\delta$
approaches zero (in this case, $\delta = 9\cdot10^{-5}$), i.e. when
input dimensions are included in the trajectories for the entropy
estimation, $\mu$ clearly converges to a plateau value. In contrast,
for $\delta=\infty$, i.e. when the input dimensions are ignored and
$\mu$ is calculated conventionally from the $x_i$ alone, convergence
to a plateau is not apparent except at almost zero perturbation
amplitude (Fig.~\ref{fig:logistic_noise}B), since it is masked by
the effect of the `unknown' input noise, which causes $\mu$ to rise
as $\epsilon \rightarrow 0$ (see \cite{faure:1998} for discussion).
Thus using the input-output method can use process noise when it is
actually known, i.e. when it is `input', to expose the
low-dimensional dynamics of the driven process. In a regime which is
periodic in the noise-free case ($a=3$.
Fig.~\ref{fig:logistic_noise}C), $\mu$ is greatly reduced. The
residual value reflects the finite number of embedding points used
and decreases with increasing N. Note however, the driving input
noise samples parts of the state space which are off the attractor,
and in general, $\mu$ is not expected to be the same as that of the
noise-free case. For example, with driving noise, a nonlinear system
can spend large amounts of time near chaotic repellors in the phase
space \cite{rand:1991}. In the synapse the `unperturbed' or
unstimulated dynamics is trivially a fixed point of amplitude zero.

\section{Synaptic transmission data}
We carried out whole cell patch-clamp recordings in 21 pairs of
synaptically-connected layer 2/3 pyramidal neurons
(Fig.~\ref{fig:ensemble_data}B), using standard techniques
\cite{methods}. The amplitude of synaptic events showed a typical
pattern of variability in repeated responses to short bursts, and
short-term depression in the ensemble average
(Fig.~\ref{fig:ensemble_data}A). Next, presynaptic APs were
stimulated continuously for periods of 30 minutes, with presynaptic
spike timing determined by an inhomogeneous Poisson process
\cite{cox} the rate of which was modulated in exponentially-decaying
bursts (peak amplitude $R_p$, time constant $\tau_b$), at times
generated by a stationary Poisson process of rate $\lambda_b$
(Fig.~\ref{fig:ensemble_data}C). Such a process is thought to model
the statistics of natural bursting synaptic input reasonably well,
and can have a coefficient of variation of interspike intervals
[CV(ISI)] greater than 1 \cite{harsch:2000}. Postsynaptic responses
to this stimulus train showed a high variability in amplitude,
including a large proportion of failures, as well as asynchronous
spontaneous events (Fig.~\ref{fig:ensemble_data}C). Distributions of
$\mathbf{\Delta t}$ and $\mathbf{A}$ are shown in
Fig.~\ref{fig:ensemble_data}D and Fig.~\ref{fig:ensemble_data}E.
Over 30 minutes of continuous stimulation at an average rate of 1.2
Hz, there is typically a small depressing trend in the average
response amplitude, referred to as long-term depression
\cite{synapsechapter9}.

\subsection{Dispersion of future synaptic transmissions}
\begin{figure}
 \begin{center}
 \begin{minipage}[b]{0.05cm}
    \textmd{A}
 \vspace{5.5cm}
 \end{minipage}
 \begin{minipage}[t]{0.95\linewidth}
 \hspace{0.1mm}
\epsfig{file=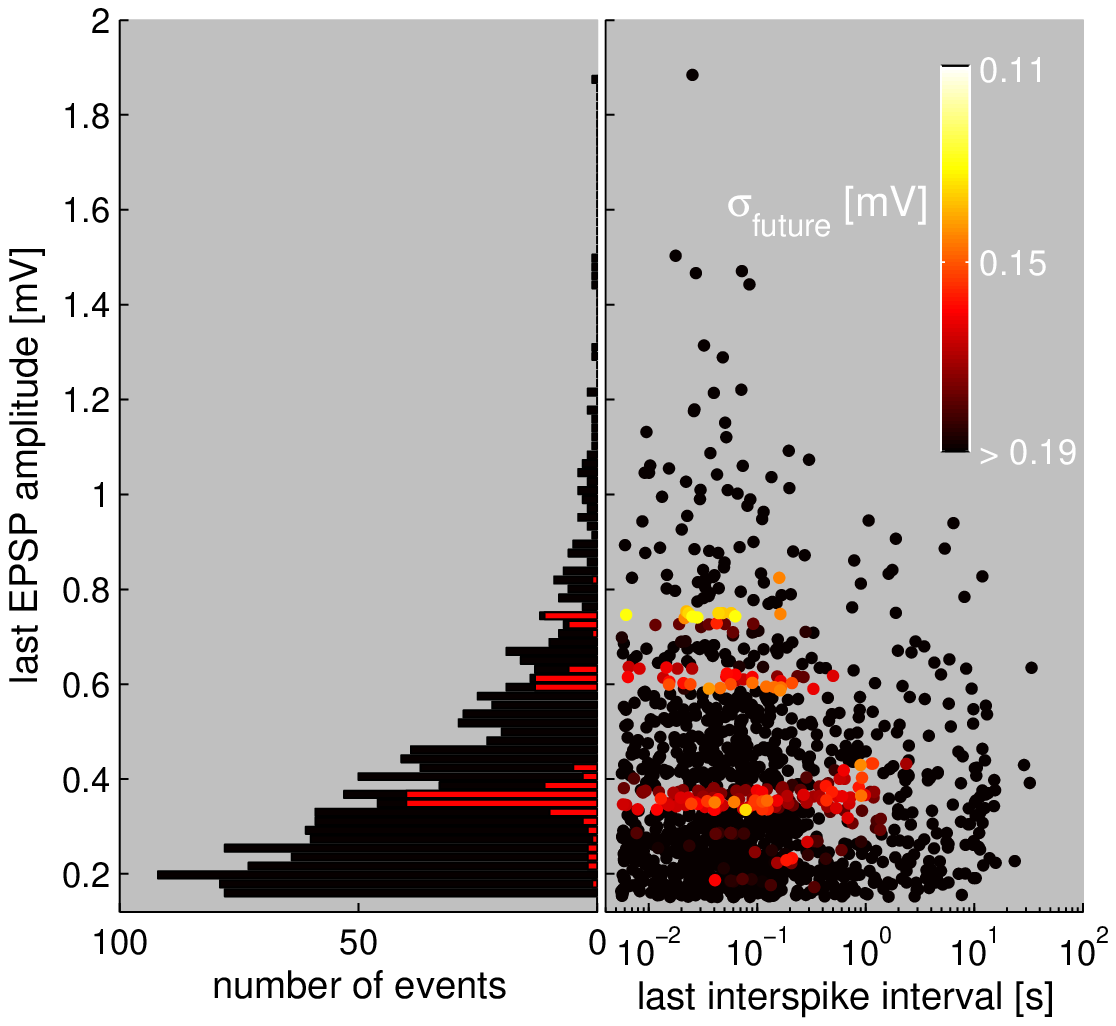,width=0.95\linewidth}
 \end{minipage}\\
 \begin{minipage}[b]{0.05cm}
 \textmd{B}
 \vspace{5.5cm}
 \end{minipage}
 \begin{minipage}[t]{0.95\linewidth}
\epsfig{file=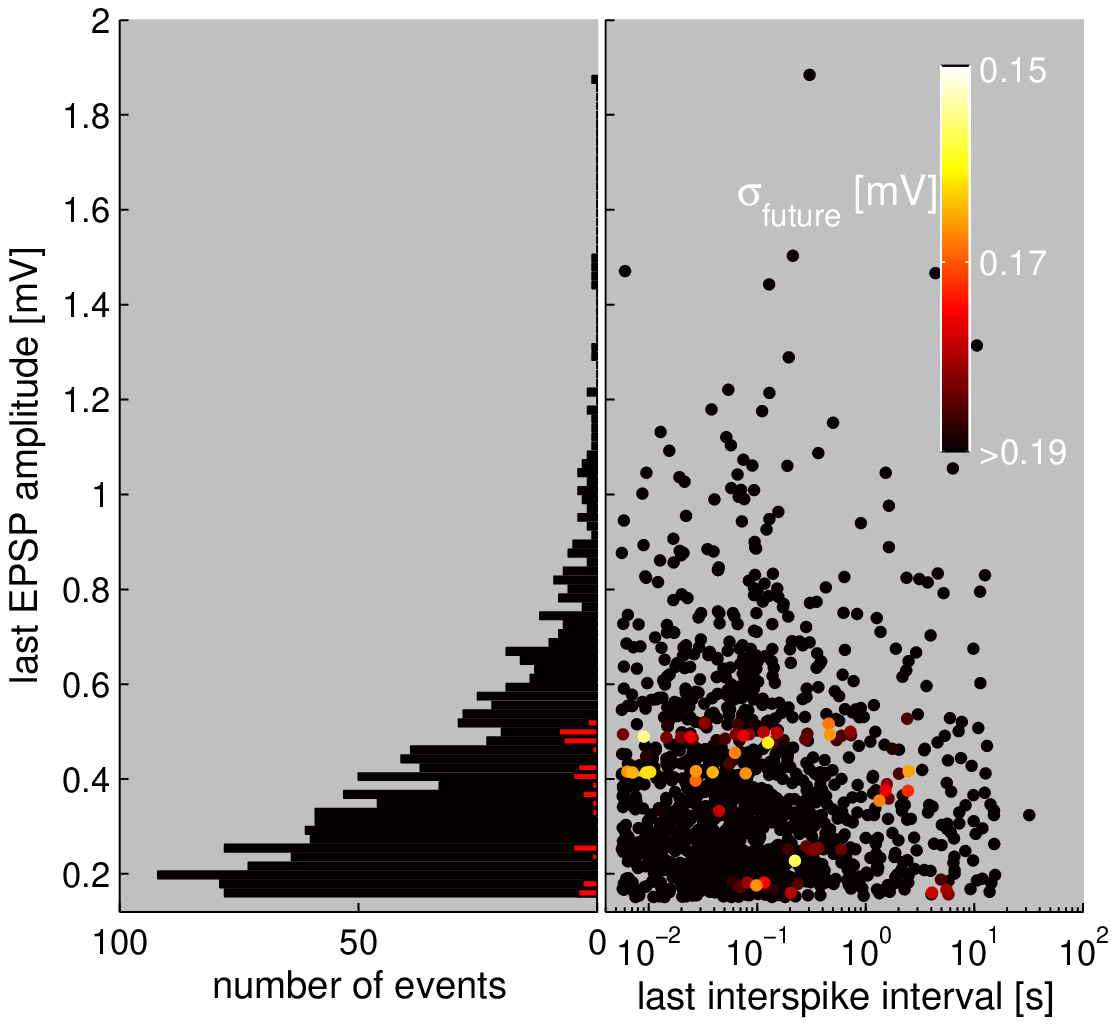,width=0.95\linewidth}
 \end{minipage}
 \end{center}
 \caption[(Color) Clustering of amplitudes of responses with reliable futures]{
 Clustering of amplitudes of responses with reliable futures.
\textbf{A}: The history vector was defined only by the last
interspike interval and last EPSP amplitude. Points are colour coded
according to the dispersion value (see text) as indicated on the
right using 50 nearest neighbours. Stimulus parameters were
$R_{peak} = 30\,\rm Hz$, $\tau_{b}=200\,\rm ms$ $\lambda_b=0.2\,\rm
Hz$, which amount to an average rate of $1.2\,\rm Hz$. \textbf{B}:
Control for A using a random permutation of the amplitude time
series.} \label{fig:lines1}
\end{figure}
First, we show evidence of nonlinear structure in the highly
variable input--output relationship of these synapses. To do this we
searched for histories $\mathbf{h_i}$  with the most reliable
futures $A_{i+1}$ (see Eqn.\ref{eqn:historyvector}). As a measure of
the similarity between two histories, we used the Euclidian distance
between their history-vectors after normalizing amplitude and
interspike interval dimensions. For each event of the time series we
computed $\mathcal{N}_k({\bf h_i})$, the set of the $k$ nearest
neighbors of ${\bf h_i}$\cite{opentstool}. The dispersion of the
future output for a trajectory $\bf h_i$ can be characterized by
$\sigma_i$, the standard deviation of the futures
$\mathcal{N}_k({\bf h_i})$ and $\sigma_A$ the standard deviation of
$\bf A$. A simple example of this analysis, which can be easily
visualized, is to distinguish which combinations of most recent
interspike interval and EPSP amplitude lead to small dispersions.
The results are presented in Fig.~\ref{fig:lines1}A. In the right
hand panel the distribution of all points in this space is shown
(the range of the axes excludes failures). Points are colored
according to their dispersions. Points with relatively reliable
futures (low dispersion) are yellow or red. It is striking that the
amplitude dimension of such points has a much tighter distribution
than that of all amplitudes (see histogram in the left panel), and
that values are clearly restricted to a few sharp peaks. This
pattern was seen in 7 out of 8 synapses analyzed in this way.
Shuffled surrogates (see Fig.~\ref{fig:lines1}B) never showed a
similar pattern. Thus, within the complex and variable response of
the synapse, a subset of patterns are transmitted with considerable
precision. In this case for intervals less than the vesicle
replenishment time constant of the synapse \cite{markram:1996,
varela:1997}, the amplitude of the preceding event has a greater
impact than the interval. The $K_2$ entropy that we have defined above gives a global
characterization of dispersion over all histories of
input--output.

\subsection{Correlation entropy of synaptic data}
\begin{figure}
\begin{center}
\epsfig{file=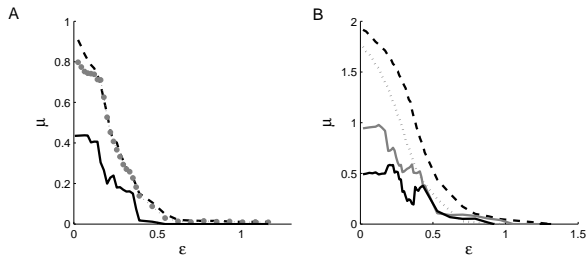,width=\linewidth}
\end{center}
\caption{Convergence of $K_2$-entropy estimate $\mu$ for synaptic
data. {\bf A}: A typical $\mu(\epsilon)$ relationship at a very
small $\delta=0.018$ms. A clear convergence is seen as $\epsilon
\rightarrow 0$. Stimulation parameters as in Fig. 1. Total number of
stimulation events in the time series was 2594. The mean of 50
randomly shuffled controls is shown as a dashed line. The standard
deviation $\sigma$ of these controls was maximal for $\epsilon
\rightarrow 0$, $\sigma<0.18$. Gray $\circ$ denotes the mean of 50
surrogates with a large shift between the amplitude and interspike
interval time series (see text). A shift $> 10$ minutes was randomly
chosen, $\sigma<0.43$ {\bf B}: $\mu(\epsilon)$ before and after
induction of spike-timing dependent plasticity (STDP). Gray line
denotes the data before the induction of STDP, black solid line
denotes after induction. The means of 50 randomly shuffled controls
are shown as corresponding dashed lines, $\sigma<0.44$ before STDP
and $\sigma<0.43$ after STDP. Stimulation parameters were $\bar{R}=
30$Hz $\lambda_{b}=0.5$Hz $\tau_{b}=400$ms, average rate 6 Hz. The
stimulation before and after STDP induction was identical, with 3000
stimulation events.} \label{fig:results}
\end{figure}

Fig.~\ref{fig:results}A shows a typical portrait of the dependence
of $\mu$ on the neighborhood size, $\epsilon$. As $\epsilon
\rightarrow 0$, $\mu$ converges to a constant plateau value (solid
black line). In surrogates where the output values are randomly
shuffled, to destroy all correlations in the output, or shifted in
time to destroy only the correspondence between the input and
output, while preserving the correlations within the input and
output individually, $\mu$ continues to grow as $\epsilon
\rightarrow 0$. Theoretically this should approach $\infty$, but is
prevented from doing so by the large number of zeros (failures) in
the amplitude distribution (Fig.~\ref{fig:ensemble_data}E). The
small difference between the two surrogates described above
indicates that there is little additional correlation in the output
which is independent from the input, i.e. the synapse is highly
driven, in this case.
Thus, a converging value for $\mu$ for small $\epsilon$ is clearly
identifiable in cortical synapses, even for stochastic input
patterns, and is a measure of the uncertainty produced by synaptic
transmission.

 $\mu$, being a measure of the information rate of
the input-output dynamics of the synapse, ought to change when the
properties of the synapse are altered via long-term plasticity,
which is believed to underlie learning and memory. To test this, we
applied a presynaptic--postsynaptic paired stimulus protocol
\cite{markram:1997,bi:2001}, in which pre and postsynaptic APs are
repeatedly stimulated with a 10 ms delay between them to allow
coincident arrival of both at the synaptic terminal. After this
so-called `spike-timing-dependent' long-term potentiation, $\mu$
showed a similar form to the control distribution, but a clear shift
to lower values in the low $\epsilon$ limit
(Fig.~\ref{fig:results}B). In this sense, less uncertainty is being
created by the synaptic transmission for the same statistics of
input, i.e. reliability of transmission is enhanced for this
stimulus process. Similar findings were seen in 5 synaptic
connections.

\subsection{A biophysical model of short-term plasticity}
To gain insight into possible underlying mechanisms, we also carried
out the same analysis on a stochastic biophysically-based
microscopic model of cortical synapses adapted from
\cite{fuhrmann:2002}. In this model, which is consistent with a
mean-field deterministic model of short-term plasticity
\cite{tsodyks:1997}, the stochastic release and replenishment of a
small pool of transmitter vesicles (`quanta') is simulated
explicitly, with Gaussian variability of vesicle amplitude. The
synaptic connection is composed of $N$ release sites. At each site
there may be, at most, one vesicle available for release, and the
release from each of the sites is independent of the release from
all other sites. The dynamics is characterized by two probabilistic
processes, release and recovery. At the arrival of a presynaptic
spike at time $ts$, each site containing a vesicle will release the
vesicle with the same probability, $U_{se}$ (use of synaptic
efficacy). Once a release occurs, the site can be refilled
(recovered) during a time interval $\Delta t$ with a probability
$1-e^{-\Delta t/\tau_{rec}}$, with $\tau_{rec}$ as a recovery time
constant. Both processes can be described by a single differential
equation, which determines the \emph{ensemble} probability, $P_{v}$,
for a vesicle to be available for release at any time $t$ \cite{fuhrmann:2002}.
\begin{equation}
\frac{dP_v}{dt}=\frac{1-P_v}{\tau_{rec}}-U_{se} \cdot P_v \cdot
\delta(t-ts)\label{eqn:diffgl}
\end{equation}
The iterative solution for a train of spikes arriving at one release site is
given by
\begin{align}
P_{v}(ts_{i+1})=&P_{v}(ts_i)\cdot(1-U_{se})\nonumber\\
\times \,& e^{-(ts_{i+1}-ts_i)/\tau_{rec}}+1-e^{-(ts_{i+1}-ts_i)/\tau_{rec}} \label{eqn:pv}
\end{align}
where $ts_i$ is the $i^{th}$ spike time, and $P_r(ts_i)=U_{se} \cdot P_v(ts_i)$ denotes the probability of release for each release
site at the time of a spike $ts_i$.\\
However, \emph{individual trajectories} follow a different dynamics
from the ensemble since at each spike time $ts$ an all-or-nothing
stochastic decision is made on the availability of a vesicle, and
therefore the release probability set back to zero if release
occurs. Thus individual realizations follow an abruptly changing,
nonlinear stochastic map. Individual trajectories were simulated as
follows: for each individual release site, the times of the most
recent transmitter release ($tr_i$) are recursively defined for a
given input time series of spike times ($ts_i$)

\begin{align}
tr_{i+1}=&
tr_i+(ts_{i+1}-tr_i)\nonumber\\
\times \, & \Theta(ts_{i+1}-tr_i-\tau_i)\Theta(U_{se}-\xi_{i+1})
\end{align}
with the associated refilling intervals:
\begin{equation}
\tau_{i+1}=\Pi_i \tilde{\Theta}(tr_{i+1}-tr_i)+\tau_i
\Theta(tr_i-tr_{i+1})
\end{equation}
where $\Pi_i$ is a random number from an exponential distribution
with a time constant $\tau_{rec}$ (see \ref{eqn:diffgl}), $\xi_i$ a
random number from a uniform distribution in the interval $[0;1]$,
and
\begin{center}
\begin{minipage}[t]{0.4\linewidth}
\begin{equation}
\Theta(x)=\left\{\begin{array}{cc} 0 & x<0 \nonumber\\
1 & x\geq 0\end{array}, \right.
\end{equation}
\end{minipage}
\begin{minipage}[t]{0.4\linewidth}
\begin{equation}
\tilde{\Theta}(x)=\left\{\begin{array}{cc} 0 & x\leq0 \nonumber\\
1 & x>0\end{array} \right.
\end{equation}
\end{minipage}
\end{center}
\vspace{12pt} The postsynaptic response $a_i$ to a single vesicle
release is assumed not to be a constant value but randomly drawn
from a Gaussian distribution characterized by the coefficient of
variation of the quantal content $CV(q)=\sigma_q / m_q$.
\begin{equation}
a_i=\zeta_i \sigma_q+m_q
\end{equation}
where $\zeta_i$ is a random number from Gaussian distribution with
mean 0 and a standard deviation of 1, and $\sigma_q$ is the standard
deviation and $m_q$ the mean of the quantal content. Negative values
for $a_i$ were rounded to zero. The final size of a postsynaptic
response $A_i$ is the sum over all $N$ release sites

\begin{align} A_{i+1}=\sum_{j=1}^{N}
a_i^j \tilde{\Theta}(tr_{i+1}^j-tr_{i}^j)
\end{align}

 In summary the model depends upon three
stochastic processes, the recovery time for vesicles $\tau_i$, the
decision whether to release $\Theta(U_{se}-\xi_{i})$, and the amount
of transmitter released $a_i$.

\begin{figure}
\begin{center}
\epsfig{file=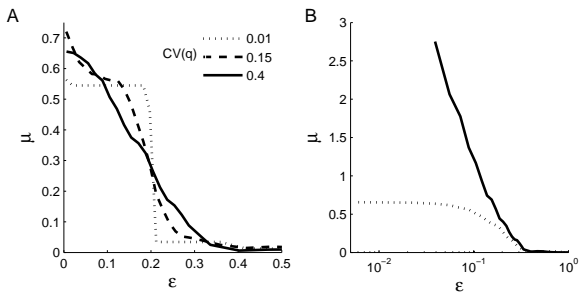,width=\linewidth}
\end{center}
\caption{{\bf A}: Analysis of a biophysically-based model of
synaptic transmission \cite{fuhrmann:2002}, driven by the same
stimulus timing as used in Fig.~\ref{fig:results}A. The model
parameters were: $U_{se}=0.5$, $\tau_{rec}=800$ ms, $N=5$, the
mean quantal size $q=0.2$ mV for all release sites, see
\cite{fuhrmann:2002} for details. Graphs are shown for three
different levels of variability of the quantal content. {\bf B}:
Same analysis as in A but with a logarithmic scale in $\epsilon$
to demonstrate the divergence between the macroscopic mean-field
model with additional extrinsic Gaussian noise (solid line) and
the microscopic stochastic model (dotted line). The standard
deviation of the Gaussian noise was estimated from average
ensemble fluctuations of 1000 surrogates of the stochastic model
(parameters as in A). The model parameters for the deterministic
model were chosen accordingly: $U_{se}=0.5$, $A_{se}=1$,
$\tau_{rec}=800$ms.} \label{fig:model}
\end{figure}
This model was able to reproduce a convergence of $\mu(\epsilon)$
for physiologically realistic parameters (Fig.~\ref{fig:model}A),
although not as flat as the experimental data. When quantal
variability was reduced to very low values, step-like patterns were
observed in the $\mu(\epsilon)$ relationship, reflecting a reliable
quantal representation (i.e. number of quanta) of EPSP amplitude.
This indicates that the convergence of $\mu$ for real data might
reflect deterministic predictability of quantal number. When we
analyzed the corresponding macroscopic or mean-field model
\cite{tsodyks:1997} with additive Gaussian noise of the same
amplitude as the average ensemble fluctuations, a very different
pattern of $\mu(\epsilon)$ was observed (Fig.~\ref{fig:model}B).
Instead of convergence at low $\epsilon$, $\mu$ rose sharply to
arbitrarily high values as $\epsilon \rightarrow 0$. Thus the
signature of this uncorrelated extrinsic noise added to the
mean-field dynamics is quite different again from what is observed
in the experimental data. The convergence of $\mu$ both in actual
data and in the microscopic biophysical model, but not for the
mean-field case with extrinsic noise, implies that the intrinsic
microscopic nature of synaptic transmission leads effectively to a
state of low-dimensional chaos.\\

Any classification of the nature of the dynamics in this way as
chaotic or stochastic, using real time series of finite length,
actually depends on the scale of $\epsilon$ and $\delta$ and the
length of data available \cite{cencini:2000}. In the nervous system,
the postsynaptic cell has a limited resolution for distinguishing
the amplitude (set by intrinsic channel gating noise) and timing
(set by jitter in synaptic latencies) of individual synaptic events.
These resolution limits, or the effective granularity of
representation of amplitude and timing, will vary according to the
location of the synapse on the dendritic tree and the state of
activation of ion channels, which together determine the spatial and
temporal filtering of inputs. Thus, the scale or
$\epsilon$,$\delta$-dependence, of the correlation entropy measure
should be meaningful physiologically in understanding the generation
and flow of information in a neural circuit.\\

Previous work has characterized synaptic reliability using very
low frequency single pulse trials (e.g. \cite{feldmeyer:2002}) or
as the ensemble responses to short bursts of APs (e.g.
\cite{abbott:1997, tsodyks:1997, dobrunz:1997, dittman:2000,
gupta:2000}). In contrast, the correlation entropy $K_2$ gives a
measure of the overall nonlinear predictability of the microscopic
input-output mapping of a synapse - driven by a particular, but
arbitrary stimulus pattern, which can be calculated practically
from synaptic data.  It also has a simple information-theoretical
interpretation as the rate of information or uncertainty
production by a synapse. Therefore, it is a natural measure for
characterizing the dynamic reliability of a synapse during natural
activity. Applying it to experimental results suggests that the
microscopic characteristics of transmitter release and
postsynaptic receptor kinetics, combined with a complex
natural-like input timing, lead to a stochastic chaotic process at
individual cortical synapses, at certain scales of resolution.

\begin{acknowledgments}
We thank Michael Small and Gonzalo de Polavieja for their comments
on an earlier version of the manuscript. Supported by grants from
the BBSRC, EC and Daiwa Anglo-Japanese Foundation. ICK was
supported by the Boehringer Ingelheim Fonds.
\end{acknowledgments}
\bibliography{syndynamics}

\begin{thebibliography}{29}
\expandafter\ifx\csname natexlab\endcsname\relax\def\natexlab#1{#1}\fi
\expandafter\ifx\csname bibnamefont\endcsname\relax
  \def\bibnamefont#1{#1}\fi
\expandafter\ifx\csname bibfnamefont\endcsname\relax
  \def\bibfnamefont#1{#1}\fi
\expandafter\ifx\csname citenamefont\endcsname\relax
  \def\citenamefont#1{#1}\fi
\expandafter\ifx\csname url\endcsname\relax
  \def\url#1{\texttt{#1}}\fi
\expandafter\ifx\csname urlprefix\endcsname\relax\def\urlprefix{URL }\fi
\providecommand{\bibinfo}[2]{#2}
\providecommand{\eprint}[2][]{\url{#2}}

\bibitem[{\citenamefont{Tsodyks and Markram}(1997)}]{tsodyks:1997}
\bibinfo{author}{\bibfnamefont{M.~V.} \bibnamefont{Tsodyks}} \bibnamefont{and}
  \bibinfo{author}{\bibfnamefont{H.}~\bibnamefont{Markram}},
  \bibinfo{journal}{Proc. Natl. Acad. Sci. USA} \textbf{\bibinfo{volume}{94}},
  \bibinfo{pages}{719} (\bibinfo{year}{1997}).

\bibitem[{\citenamefont{Abbott et~al.}(1997)\citenamefont{Abbott, Varela, Sen,
  and Nelson}}]{abbott:1997}
\bibinfo{author}{\bibfnamefont{L.~F.} \bibnamefont{Abbott}},
  \bibinfo{author}{\bibfnamefont{J.~A.} \bibnamefont{Varela}},
  \bibinfo{author}{\bibfnamefont{K.}~\bibnamefont{Sen}}, \bibnamefont{and}
  \bibinfo{author}{\bibfnamefont{S.~B.} \bibnamefont{Nelson}},
  \bibinfo{journal}{Science} \textbf{\bibinfo{volume}{275}},
  \bibinfo{pages}{220} (\bibinfo{year}{1997}).

\bibitem[{\citenamefont{Varela et~al.}(1997)\citenamefont{Varela, Kamal,
  Gibson, Fost, Abbott, and Nelson}}]{varela:1997}
\bibinfo{author}{\bibfnamefont{J.~A.} \bibnamefont{Varela}},
  \bibinfo{author}{\bibfnamefont{S.}~\bibnamefont{Kamal}},
  \bibinfo{author}{\bibfnamefont{J.}~\bibnamefont{Gibson}},
  \bibinfo{author}{\bibfnamefont{J.}~\bibnamefont{Fost}},
  \bibinfo{author}{\bibfnamefont{L.~F.} \bibnamefont{Abbott}},
  \bibnamefont{and} \bibinfo{author}{\bibfnamefont{S.~B.}
  \bibnamefont{Nelson}}, \bibinfo{journal}{J. Neurosci.}
  \textbf{\bibinfo{volume}{17}}, \bibinfo{pages}{7926} (\bibinfo{year}{1997}).

\bibitem[{\citenamefont{Dobrunz and Stevens}(1997)}]{dobrunz:1997}
\bibinfo{author}{\bibfnamefont{L.~E.} \bibnamefont{Dobrunz}} \bibnamefont{and}
  \bibinfo{author}{\bibfnamefont{C.~F.} \bibnamefont{Stevens}},
  \bibinfo{journal}{Neuron} \textbf{\bibinfo{volume}{18}}, \bibinfo{pages}{995}
  (\bibinfo{year}{1997}).

\bibitem[{\citenamefont{Thomson}(1997)}]{thomson:1997b}
\bibinfo{author}{\bibfnamefont{A.}~\bibnamefont{Thomson}}, \bibinfo{journal}{J
  Physiol (Lond)} \textbf{\bibinfo{volume}{502}}, \bibinfo{pages}{131}
  (\bibinfo{year}{1997}).

\bibitem[{\citenamefont{Dobrunz and Stevens}(1999)}]{dobrunz:1999}
\bibinfo{author}{\bibfnamefont{L.~E.} \bibnamefont{Dobrunz}} \bibnamefont{and}
  \bibinfo{author}{\bibfnamefont{C.~F.} \bibnamefont{Stevens}},
  \bibinfo{journal}{Neuron} \textbf{\bibinfo{volume}{22}}, \bibinfo{pages}{157}
  (\bibinfo{year}{1999}).

\bibitem[{\citenamefont{Stevens and Wang}(1995)}]{stevens:1995}
\bibinfo{author}{\bibfnamefont{C.~F.} \bibnamefont{Stevens}} \bibnamefont{and}
  \bibinfo{author}{\bibfnamefont{Y.}~\bibnamefont{Wang}},
  \bibinfo{journal}{Neuron} \textbf{\bibinfo{volume}{14}}, \bibinfo{pages}{795}
  (\bibinfo{year}{1995}).

\bibitem[{\citenamefont{Rand and Wilson}(1991)}]{rand:1991}
\bibinfo{author}{\bibfnamefont{D.}~\bibnamefont{Rand}} \bibnamefont{and}
  \bibinfo{author}{\bibfnamefont{H.}~\bibnamefont{Wilson}},
  \bibinfo{journal}{Proc.R. Soc. Lond.B} \textbf{\bibinfo{volume}{246}},
  \bibinfo{pages}{179} (\bibinfo{year}{1991}).

\bibitem[{\citenamefont{Cao et~al.}(1998{\natexlab{a}})\citenamefont{Cao, Mees,
  and Judd}}]{cao:1998a}
\bibinfo{author}{\bibfnamefont{L.}~\bibnamefont{Cao}},
  \bibinfo{author}{\bibfnamefont{A.}~\bibnamefont{Mees}}, \bibnamefont{and}
  \bibinfo{author}{\bibfnamefont{K.}~\bibnamefont{Judd}},
  \bibinfo{journal}{Physica D} \textbf{\bibinfo{volume}{121}},
  \bibinfo{pages}{75} (\bibinfo{year}{1998}{\natexlab{a}}).

\bibitem[{\citenamefont{Cao et~al.}(1998{\natexlab{b}})\citenamefont{Cao, Mees,
  Judd, and Froyland}}]{cao:1998b}
\bibinfo{author}{\bibfnamefont{L.}~\bibnamefont{Cao}},
  \bibinfo{author}{\bibfnamefont{A.~I.} \bibnamefont{Mees}},
  \bibinfo{author}{\bibfnamefont{K.}~\bibnamefont{Judd}}, \bibnamefont{and}
  \bibinfo{author}{\bibfnamefont{G.}~\bibnamefont{Froyland}},
  \bibinfo{journal}{Int. J. Bifurcation and Chaos}
  \textbf{\bibinfo{volume}{8}}, \bibinfo{pages}{1491}
  (\bibinfo{year}{1998}{\natexlab{b}}).

\bibitem[{\citenamefont{Grassberger and Procaccia}(1983)}]{grassberger:1983}
\bibinfo{author}{\bibfnamefont{P.}~\bibnamefont{Grassberger}} \bibnamefont{and}
  \bibinfo{author}{\bibfnamefont{I.}~\bibnamefont{Procaccia}},
  \bibinfo{journal}{Phys.Rev.A} \textbf{\bibinfo{volume}{28}},
  \bibinfo{pages}{2591} (\bibinfo{year}{1983}).

\bibitem[{\citenamefont{Gaspard and Wang}(1993)}]{gaspard:1993}
\bibinfo{author}{\bibfnamefont{P.}~\bibnamefont{Gaspard}} \bibnamefont{and}
  \bibinfo{author}{\bibfnamefont{X.-J.} \bibnamefont{Wang}},
  \bibinfo{journal}{Phys.Rep.} \textbf{\bibinfo{volume}{235}},
  \bibinfo{pages}{291} (\bibinfo{year}{1993}).

\bibitem[{\citenamefont{Schittenkopf and Deco}(1997)}]{schittenkopf:1997}
\bibinfo{author}{\bibfnamefont{C.}~\bibnamefont{Schittenkopf}}
  \bibnamefont{and} \bibinfo{author}{\bibfnamefont{G.}~\bibnamefont{Deco}},
  \bibinfo{journal}{PHYSICA D} \textbf{\bibinfo{volume}{110}},
  \bibinfo{pages}{173} (\bibinfo{year}{1997}).

\bibitem[{\citenamefont{Faure and Korn}(1998)}]{faure:1998}
\bibinfo{author}{\bibfnamefont{P.}~\bibnamefont{Faure}} \bibnamefont{and}
  \bibinfo{author}{\bibfnamefont{H.}~\bibnamefont{Korn}},
  \bibinfo{journal}{Physica D} \textbf{\bibinfo{volume}{122}},
  \bibinfo{pages}{265} (\bibinfo{year}{1998}).

\bibitem[{\citenamefont{{Webber Jr.} and Zbilut}(1994)}]{webber:1994}
\bibinfo{author}{\bibfnamefont{C.}~\bibnamefont{{Webber Jr.}}}
  \bibnamefont{and} \bibinfo{author}{\bibfnamefont{J.}~\bibnamefont{Zbilut}},
  \bibinfo{journal}{J.Appl.Physiol.} \textbf{\bibinfo{volume}{76}},
  \bibinfo{pages}{965} (\bibinfo{year}{1994}).

\bibitem[{met()}]{methods}
\emph{\bibinfo{title}{{\emph{\scriptsize Slices of 300$\mu$m thickness from the
  somatosensory cortex of Wistar rats (12-21 days old) were cut using a
  vibrating microslicer (Campden Instruments, UK) in cold (0-4$^{\circ}$C)
  oxygenated artificial cerebrospinal fluid solution containing (in [mM]): 125
  NaCl, 2.5 KCL, 25 NaHCO$_3$, 25 glucose, 1.25 NaH$_2$PO$_4$, 2 CaCl$_2$, 1
  MgCl$_2$. After the slicing procedure, the slices were kept at room
  temperature. The intracellular solution containing (in [mM])105 Potassium
  Gluconate, 30 KCL, 10 HEPES, 10 Phosphocreatine-Na$_2$, 4 ATP-Mg, 0.3 GTP was
  adjusted to pH 7.3 with KOH. Multiple patch-clamp recordings in the whole
  cell configuration were carried out using combinations of up to four
  patch-clamp amplifiers (Axon Instruments, USA) to maximize the probability of
  synaptic connections. The data were amplified, filtered (5kHz Low Pass
  Bessel), digitized and sampled at 20kHz (see also \cite{stuart:1995})}}}}.

\bibitem[{\citenamefont{Cox and Miller}(1977)}]{cox}
\bibinfo{author}{\bibfnamefont{D.}~\bibnamefont{Cox}} \bibnamefont{and}
  \bibinfo{author}{\bibfnamefont{H.}~\bibnamefont{Miller}},
  \emph{\bibinfo{title}{The Theory of Stochastic Processes}}
  (\bibinfo{publisher}{Chapman \& Hall/CRC}, \bibinfo{year}{1977}).

\bibitem[{\citenamefont{Harsch and Robinson}(2000)}]{harsch:2000}
\bibinfo{author}{\bibfnamefont{A.}~\bibnamefont{Harsch}} \bibnamefont{and}
  \bibinfo{author}{\bibfnamefont{H.~P.~C.} \bibnamefont{Robinson}},
  \bibinfo{journal}{J. Neurosci.} \textbf{\bibinfo{volume}{20}},
  \bibinfo{pages}{6181} (\bibinfo{year}{2000}).

\bibitem[{\citenamefont{Malenka and Siegelbaum}(2001)}]{synapsechapter9}
\bibinfo{author}{\bibfnamefont{R.~C.} \bibnamefont{Malenka}} \bibnamefont{and}
  \bibinfo{author}{\bibfnamefont{S.~A.} \bibnamefont{Siegelbaum}},
  \emph{\bibinfo{title}{Synapses}} (\bibinfo{publisher}{John Hopkins University
  Press}, \bibinfo{address}{Baltimore, Maryland}, \bibinfo{year}{2001}),
  chap.~\bibinfo{chapter}{9}, pp. \bibinfo{pages}{393--453}.

\bibitem[{ope()}]{opentstool}
\emph{\bibinfo{title}{{\emph{\scriptsize We used a box-assisted algorithm for
  the nearest neighbor search, exluding points with overlapping histories, from
  OpenTSTOOL, Version 1.11 by Merkwirth, Parlitz, Wedekind and Lauterborn, DPI,
  G{\"o}ttingen, Germany}}}}.

\bibitem[{\citenamefont{Markram and Tsodyks}(1996)}]{markram:1996}
\bibinfo{author}{\bibfnamefont{H.}~\bibnamefont{Markram}} \bibnamefont{and}
  \bibinfo{author}{\bibfnamefont{M.}~\bibnamefont{Tsodyks}},
  \bibinfo{journal}{Nature} \textbf{\bibinfo{volume}{382}},
  \bibinfo{pages}{807} (\bibinfo{year}{1996}).

\bibitem[{\citenamefont{Markram et~al.}(1997)\citenamefont{Markram, L\"{u}bke,
  Frotscher, and Sakmann}}]{markram:1997}
\bibinfo{author}{\bibfnamefont{H.}~\bibnamefont{Markram}},
  \bibinfo{author}{\bibfnamefont{J.}~\bibnamefont{L\"{u}bke}},
  \bibinfo{author}{\bibfnamefont{M.}~\bibnamefont{Frotscher}},
  \bibnamefont{and} \bibinfo{author}{\bibfnamefont{B.}~\bibnamefont{Sakmann}},
  \bibinfo{journal}{Science} \textbf{\bibinfo{volume}{275}},
  \bibinfo{pages}{213} (\bibinfo{year}{1997}).

\bibitem[{\citenamefont{Bi and Poo}(2001)}]{bi:2001}
\bibinfo{author}{\bibfnamefont{G.~Q.} \bibnamefont{Bi}} \bibnamefont{and}
  \bibinfo{author}{\bibfnamefont{M.~M.} \bibnamefont{Poo}},
  \bibinfo{journal}{Annu. Rev. Neurosci.} \textbf{\bibinfo{volume}{24}},
  \bibinfo{pages}{139} (\bibinfo{year}{2001}).

\bibitem[{\citenamefont{Fuhrmann et~al.}(2002)\citenamefont{Fuhrmann, Segev,
  Markram, and Tsodyks}}]{fuhrmann:2002}
\bibinfo{author}{\bibfnamefont{G.}~\bibnamefont{Fuhrmann}},
  \bibinfo{author}{\bibfnamefont{I.}~\bibnamefont{Segev}},
  \bibinfo{author}{\bibfnamefont{H.}~\bibnamefont{Markram}}, \bibnamefont{and}
  \bibinfo{author}{\bibfnamefont{M.}~\bibnamefont{Tsodyks}},
  \bibinfo{journal}{J. Neurophysiol.} \textbf{\bibinfo{volume}{87}},
  \bibinfo{pages}{140} (\bibinfo{year}{2002}).

\bibitem[{\citenamefont{Cencini et~al.}(2000)\citenamefont{Cencini, Falcioni,
  Olbrich, Kantz, and Vulpiani}}]{cencini:2000}
\bibinfo{author}{\bibfnamefont{M.}~\bibnamefont{Cencini}},
  \bibinfo{author}{\bibfnamefont{M.}~\bibnamefont{Falcioni}},
  \bibinfo{author}{\bibfnamefont{E.}~\bibnamefont{Olbrich}},
  \bibinfo{author}{\bibfnamefont{H.}~\bibnamefont{Kantz}}, \bibnamefont{and}
  \bibinfo{author}{\bibfnamefont{A.}~\bibnamefont{Vulpiani}},
  \bibinfo{journal}{Phys.Rev.E} \textbf{\bibinfo{volume}{62}},
  \bibinfo{pages}{427} (\bibinfo{year}{2000}).

\bibitem[{\citenamefont{Feldmeyer et~al.}(2002)\citenamefont{Feldmeyer, Lubke,
  Silver, and Sakmann}}]{feldmeyer:2002}
\bibinfo{author}{\bibfnamefont{D.}~\bibnamefont{Feldmeyer}},
  \bibinfo{author}{\bibfnamefont{J.}~\bibnamefont{Lubke}},
  \bibinfo{author}{\bibfnamefont{R.~A.} \bibnamefont{Silver}},
  \bibnamefont{and} \bibinfo{author}{\bibfnamefont{B.}~\bibnamefont{Sakmann}},
  \bibinfo{journal}{J Physiol (Lond)} \textbf{\bibinfo{volume}{538}},
  \bibinfo{pages}{803} (\bibinfo{year}{2002}).

\bibitem[{\citenamefont{Dittman et~al.}(2000)\citenamefont{Dittman, Kreitzer,
  and Regehr}}]{dittman:2000}
\bibinfo{author}{\bibfnamefont{J.~S.} \bibnamefont{Dittman}},
  \bibinfo{author}{\bibfnamefont{A.~C.} \bibnamefont{Kreitzer}},
  \bibnamefont{and} \bibinfo{author}{\bibfnamefont{W.~G.}
  \bibnamefont{Regehr}}, \bibinfo{journal}{J. Neurosci.}
  \textbf{\bibinfo{volume}{20}}, \bibinfo{pages}{1374} (\bibinfo{year}{2000}).

\bibitem[{\citenamefont{Gupta et~al.}(2000)\citenamefont{Gupta, Wang, and
  Markram}}]{gupta:2000}
\bibinfo{author}{\bibfnamefont{A.}~\bibnamefont{Gupta}},
  \bibinfo{author}{\bibfnamefont{Y.}~\bibnamefont{Wang}}, \bibnamefont{and}
  \bibinfo{author}{\bibfnamefont{H.}~\bibnamefont{Markram}},
  \bibinfo{journal}{Science} \textbf{\bibinfo{volume}{287}},
  \bibinfo{pages}{273} (\bibinfo{year}{2000}).

\bibitem[{\citenamefont{Sakmann and Stuart}(1995)}]{stuart:1995}
\bibinfo{author}{\bibfnamefont{B.}~\bibnamefont{Sakmann}} \bibnamefont{and}
  \bibinfo{author}{\bibfnamefont{G.}~\bibnamefont{Stuart}},
  \emph{\bibinfo{title}{Single Channel Recording}} (\bibinfo{publisher}{Plenum
  Press}, \bibinfo{address}{New York}, \bibinfo{year}{1995}), chap.
  \bibinfo{chapter}{8. Patch-clamp recordings from the soma, dendrites and axon
  of neurons in brain slices}, pp. \bibinfo{pages}{199--212},
  \bibinfo{edition}{2nd} ed.

\end{thebibliography}

\end{document}